\documentclass[reprint,amssymb, amsmath, aps, superscriptaddress, showpacs, prb]{revtex4-1}
\usepackage{graphicx}
\usepackage{hyperref}
\usepackage{xcolor}
\begin{document}

\title{Mo$_{6}$Ga$_{31}$ endohedral cluster superconductor}
\author{Valeriy Yu. \surname{Verchenko}}
\affiliation{Department of Chemistry, Lomonosov Moscow State University, 119991 Moscow, Russia}
\affiliation{National Institute of Chemical Physics and Biophysics, 12618 Tallinn, Estonia}
\email{valeriy.verchenko@gmail.com}

\author{Alexander O. \surname{Zubtsovskii}}
\affiliation{Department of Chemistry, Lomonosov Moscow State University, 119991 Moscow, Russia}
\affiliation{Experimental Physics VI, Center for Electronic Correlations and Magnetism, Institute of Physics, University of Augsburg, 86135 Augsburg, Germany}

\author{Alexander A. \surname{Tsirlin}}
\affiliation{Experimental Physics VI, Center for Electronic Correlations and Magnetism, Institute of Physics, University of Augsburg, 86135 Augsburg, Germany}

\author{Zheng \surname{Wei}}
\affiliation{Department of Chemistry, University at Albany, SUNY, Albany, 12222 New York, United States}

\author{Maria \surname{Roslova}}
\affiliation{Department of Materials and Environmental Chemistry, Stockholm University, SE-106 91 Stockholm, Sweden}

\author{Evgeny V. \surname{Dikarev}}
\affiliation{Department of Chemistry, University at Albany, SUNY, Albany, 12222 New York, United States}

\author{Andrei V. \surname{Shevelkov}}
\affiliation{Department of Chemistry, Lomonosov Moscow State University, 119991 Moscow, Russia}

\begin{abstract}
Synthesis, crystal and electronic structure, and physical properties of the Mo$_6$Ga$_{31}$ endohedral cluster superconductor are reported. The compound has two crystallographic modifications, monoclinic and triclinic, in which the same \{Mo$_{12}$Ga$_{62}$\} building units are perpendicular or codirectional to each other, respectively. Monoclinic and triclinic structures of Mo$_6$Ga$_{31}$ possess qualitatively the same electronic density of states showing a high peak at the Fermi level. Both modifications are inherently present in the bulk specimen. Due to the proximity effect, bulk Mo$_6$Ga$_{31}$ exhibits single superconducting transition at the critical temperature of 8.2\,K in zero magnetic field. The upper critical field, which is 7.8\,T at zero temperature, shows clear enhancement with respect to the Werthamer-Helfand-Honenberg prediction. Accordingly, heat capacity measurements indicate strong electron-phonon coupling in the superconducting state with the large ratio of $2\Delta(0)/(k_BT_c)=4.5$, where 2$\Delta(0)$ is the full superconducting gap at zero temperature.
\end{abstract}

\maketitle

\section{Introduction}

Recently, the transition metal-embedded Ga clusters were proposed as a structural motif favorable for superconductivity.\cite{rega5} Endohedral Ga clusters centered by atoms of 4$d$ or 5$d$ transition metals can be found in the crystal structures of Ga-rich binary intermetallic compounds, among which Mo$_{8}$Ga$_{41}$ with $T_c=9.8$\,K,\cite{mo8ga41-1} Mo$_{6}$Ga$_{31}$ (8\,K),\cite{mo6ga31-1} Mo$_4$Ga$_{21-x-\delta}$Sn$_x$ (5.85\,K),\cite{mogasn} ReGa$_5$ (2.3\,K),\cite{rega5} Rh$_2$Ga$_9$ (1.9\,K),\cite{t2ga9} and Ir$_2$Ga$_9$ (2.2\,K)\cite{t2ga9} exhibit superconducting properties. In this list, the Mo-based superconductors are distinguished by higher critical temperatures. A closer look reveals that they possess nontrivial superconducting-state propeties deviating from the Bardeen-Cooper-Schriffer (BCS) model.

For Mo$_8$Ga$_{41}$, measurements of heat capacity indicate strong coupling superconductivity with the full superconducting gap of $2\Delta(0)/(k_BT_c)=4.4$ exceeding significantly the weak-coupling BCS limit.\cite{mo8ga41-2,mo8ga41-3} Furthermore, muon spin rotation/relaxation ($\mu$SR) experiments show possible multigap or multiband superconductivity\cite{mo8ga41-4} that may originate as a result of the site-selective mechanism involving two Fermi-surface sheets with different band velocities.\cite{mo8ga41-5} In the two independent studies, scanning tunneling spectroscopy was employed to directly probe the multigap behavior of Mo$_8$Ga$_{41}$. While measurements on a polycrystalline sample confirmed the two-gap scenario,\cite{mo8ga41-3} study of single crystals revealed the formation of surface domains, where spatially resolved single-gap order parameter was observed.\cite{mo8ga41-5}

Besides Mo$_8$Ga$_{41}$, there are Mo$_6$Ga$_{31}$\cite{mo6ga31-1} and Mo$_4$Ga$_{21-x-\delta}$Sn$_x$\cite{mogasn} endohedral cluster superconductors, and the latter shows strong coupling superconductivity similar in nature to Mo$_8$Ga$_{41}$. Focusing on Mo$_6$Ga$_{31}$ with $T_c=8$\,K, its properties are scarcely characterized in the literature.\cite{mo6ga31-1} The available information is based on the AC susceptibility and magnetization measurements carried out on a sample with the MoGa$_4$ nominal composition, which do not correspond to the actual composition of the compound. Moreover, the exact elemental and phase composition of the sample were not reported in Ref.~\cite{mo6ga31-1}. The measurements revealed superconductivity of the sample below $T_c=8$\,K in zero magnetic field with the upper critical field of $\mu_0H_{c2}(0)=7.4$\,T extrapolated to zero temperature.\cite{mo6ga31-1}

\begin{figure}
\includegraphics{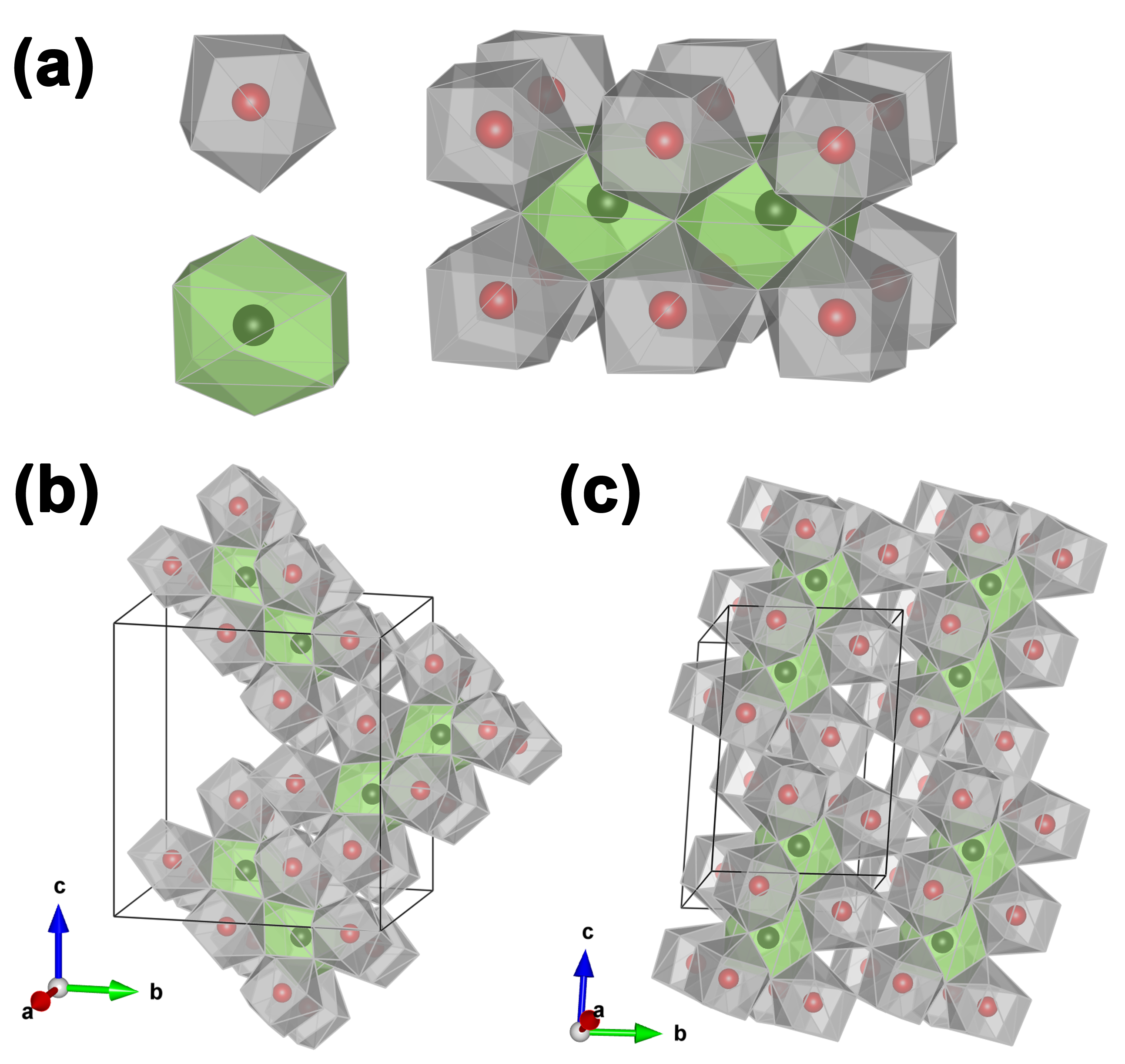}
\caption{\label{f1} (a) Main building unit of the Mo$_6$Ga$_{31}$ crystal structure composed of Mo@Ga$_{10}$ polyhedra (gray) and Ga@Ga$_{12}$ cuboctahedra (green). The arrangement of building units in the monoclinic (b) and triclinic (c) structures is shown in the bottom part of figure.}
\end{figure}

Crystal structure of Mo$_6$Ga$_{31}$ was probed in two independent studies. In the original investigation, monoclinic crystal structure, the $P2_1/c$ space group, was reported.\cite{mo6ga31-2} However, single crystals, which adopt triclinic crystal structure, the $P$-1 space group, were also obtained under different synthetic conditions.\cite{mo6ga31-3} Both structures of Mo$_6$Ga$_{31}$, monoclinic and triclinic, are based on the same building unit, which is shown in Figure~\ref{f1}(a). The main building unit consists of Mo-embedded Mo@Ga$_{10}$ clusters and Ga-centered Ga@Ga$_{12}$ cuboctahedra. Twelve Mo@Ga$_{10}$ clusters form a distorted rectangular superprism, \{Mo$_{12}$Ga$_{62}$\}, where each Mo@Ga$_{10}$ cluster shares its triangular faces with the Ga@Ga$_{12}$ cuboctahedron, and two adjacent Ga@Ga$_{12}$ cuboctahedra have a common rectangular face. In the monoclinic and triclinic structures, these building units connected by corners are perpendicular or codirectional to each other, respectively (Fig.~\ref{f1}(b,c)).

Here, we report on the synthesis, physical properties, and electronic structure of the Mo$_6$Ga$_{31}$ endohedral cluster superconductor, a compound, which is closely related to the Mo$_8$Ga$_{41}$ and Mo$_4$Ga$_{21-x-\delta}$Sn$_x$ superconductors in the strong coupling regime. Mo$_8$Ga$_{41}$, Mo$_6$Ga$_{31}$, and Mo$_4$Ga$_{21-x-\delta}$Sn$_x$ build up structural series with the general formula of Mo$_n$Ga$_{5n+1}$, where $n$ corresponds to the number of Mo@Ga$_{10}$ polyhedra (Mo@Ga$_9$Sn in the case of Mo$_4$Ga$_{21-x-\delta}$Sn$_x$) condensed on one Ga@Ga$_{12}$ cuboctahedron in the main building unit. In this study, we investigate the superconducting state of Mo$_6$Ga$_{31}$ in order to measure the electron-phonon coupling  and analyze how it correlates with the critical temperature, density of states at the Fermi level, valence electron count and other crucial parameters in the Mo$_n$Ga$_{5n+1}$ series of superconductors.

\section{Experimental details}

Mo$_6$Ga$_{31}$ was synthesized using the standard ampule technique. The stoichiometric mixture of Mo ($4N$, powder) and Ga ($5N$, pieces) was placed inside a quartz ampule, which was evacuated to the residual pressure of $5\times10^{-3}$\,mbar and flame-sealed. To obtain a polycrystalline specimen of Mo$_6$Ga$_{31}$, the ampule was annealed in a programmable furnace at 700\,$^{\circ}$C for 14~days with one intermediate grinding. The resulting black powder, which contains Mo$_6$Ga$_{31}$ with no admixture of other compounds, was used for thermodynamic and electrical transport measurements. Single crystals of Mo$_6$Ga$_{31}$ suitable for structural studies were selected from the specimens prepared under different synthetic conditions employing crystal growth from the high-temperature melt (see Section III.A for details).

High-resolution powder X-ray diffraction (HRPXRD) measurements were performed at the ID22 beam line ($\lambda=0.35451(1)$\,\r A, $2\theta_{max}=28$\,$^{\circ}$) of the European Synchrotron Radiation Facility (ESRF, Grenoble, France). Measurements were conducted at room temperature and at elevated temperatures using a hot-air blower on a sample enclosed in a fused silica capillary with a diameter of 0.3\,mm. Le Bail fittings of the HRPXRD data were performed using the Jana2006 program.\cite{jana} HRPXRD patterns collected at elevated temperatures are presented in the Supporting Information. Polycrystalline specimen of Mo$_6$Ga$_{31}$ was studied by differential scanning calorimetry (DSC) using a STA 409 PC Luxx thermal analyzer (Netzsch). Measurements were performed in high-purity Ar atmosphere at temperatures between 30\,$^{\circ}$C and 600\,$^{\circ}$C with the heating/cooling rate of 10\,$^{\circ}$C/min, and the results are presented in the Supporting Information.

Three-dimensional electron diffraction (3D ED) patterns were collected on a Themis~Z transmission electron microscope operated at 300\,kV employing the InsteaDMatic script\cite{instead} for data acquisition. In a typical experiment, a crystal is continuously rotated while ED frames are collected over the tilt range of $\pm 50$\,$^{\circ}$ with the rotation speed of 0.43\,\,$^{\circ}$/s. The exposure time of 0.3\,s was used in the experiments. 3D ED patterns were visualized by the REDp program.\cite{red} The collected 3D ED patterns and the corresponding energy-dispersive X-ray (EDX) specta are presented in the Supporting Information.

Single-crystal X-ray diffraction experiments were perfomed on a Bruker D8 Venture diffractometer (Mo X-ray source, graphite monochromator, $\lambda=0.71073$\,\r A, $T=100$\,K) equipped with a Photon 100 CMOS detector. For the absorption correction, the multi-scan routine was employed. The crystal structures were determined by the charge-flipping algorithm using the Superflip program,\cite{superflip} and refiend against $|F^2|$ using the SHELXL-2018\cite{shelx} and Jana2006 programs.\cite{jana} The atomic coordinates were standardized using the STRUCTURE TIDY program\cite{tidy} as implemented in the VESTA software,\cite{vesta} which was also used for visualization of crystal structures.

Electronic structure calculations were performed within the framework of density functional theory using the full-potential local-orbital minimum-basis band-structure code FPLO (version 14.00-47).\cite{fplo} The experimental structural data based on single-crystal XRD measurements were used in calculations.  In the scalar relativistic regime, local density aproximation was used to treat the exchange and correlation energy.\cite{lda} Integrations were performed by the improved tetrahedron method\cite{integr} on a grid of $12\times12\times12$ $k$-points in the first Brillouin zone.

Electrical resistivity was measured by the standard four-probe technique using the AC transport option of a Physical Property Measurement System (PPMS, Quantum Design) at temperatures between 2\,K and 300\,K in magnetic fields from 0\,T to 10\,T. For measurements, a rectangular-shaped pellet with typical dimensions of $8\times3\times1$\,mm$^3$ was prepared by pressing the polycrystalline specimen at the external pressure of 4\,kbar at room temperature. Cu wires with the diameter of 100\,$\mu$m were attached to the pellet using silver-containing epoxy resin. Magnetization measurements were conducted on a Magnetic Properties Measurement System (MPMS 3 SQUID, Quantum Design) in the zero-field-cooling (zfc) and field-cooling (fc) conditions in the temperature range of 1.8--15\,K in the magnetic field of 5\,Oe. Also, magnetization was measured in the zfc conditions at various fixed temperatures between 2\,K and 8\,K by sweeping magnetic field from 0\,T to 14\,T using the VSM option of PPMS. Heat capacity was measured using a relaxation-type calorimeter (HC option of PPMS, Quantum Design) at temperatures between 1.8\,K and 20\,K in magnetic fields from 0\,T to 10\,T.

\section{RESULTS AND DISCUSSION}

\subsection{Synthesis and crystal structure}

Although two crystallographic modifications of Mo$_6$Ga$_{31}$ were reported, it is unclear, at which experimental conditions they can be obtained separately. Information on synthesis of the monoclinic Mo$_6$Ga$_{31}$ is missing in the original study,\cite{mo6ga31-2} while single crystals of the triclinic Mo$_6$Ga$_{31}$ were obtained from the MoGa$_9$Si sample, where the Mo$_6$Ga$_{31}$ phase was a side product.\cite{mo6ga31-3} We systematically studied synthetic conditions, at which single crystals of Mo$_6$Ga$_{31}$ can be obtained. The use of excess of Ga metal leads to crystallization of the Mo$_8$Ga$_{41}$ phase as the main product. Therefore, we performed syntheses of samples with the stoichiometric composition, while controlling the annealing temperature and the cooling rate. The annealing at 700\,$^{\circ}$C yields polycrystalline Mo$_6$Ga$_{31}$, whereas single crystals can be obtained by increasing the annealing temperature up to 1000\,$^{\circ}$C. Single crystals of the monoclinic Mo$_6$Ga$_{31}$ were selected from the sample, which was allowed to cool down to room temperature in the shut-off furnace (fast cooling). Single crystals adopting the triclinic structure were found in the sample, which was cooled at the rate of 4\,$^{\circ}$C/h (slow cooling). In both cases, tiny submillimeter-size single crystals were isolated. The bulk specimen synthesized at 700\,$^{\circ}$C contains solely polycrystalline Mo$_6$Ga$_{31}$.

Single crystals selected from the specimens were studied by single-crystal X-ray diffraction. Tables~S1-S3 of the Supporting Information summarize the results, which are in good agreement with the previous reports,\cite{mo6ga31-2,mo6ga31-3} confirming the formation of two crystallographic modifications of Mo$_6$Ga$_{31}$. Here, we comment on the outstanding complexity of the crystal structures: the monoclinic structure contains 38 crystallographic sites [$V=2566.78(3)$\,\r A$^3$, $Z=4$, 148 atoms in the primitive cell], while there are 39 sites in the triclinic structure [$V=1282.75(2)$\,\r A$^3$, $Z=2$, 74 atoms in the primitive cell]. Given this complexity, it is natural to assume the formation of various types of defects in a polycrystalline specimen.

\begin{figure}
\includegraphics{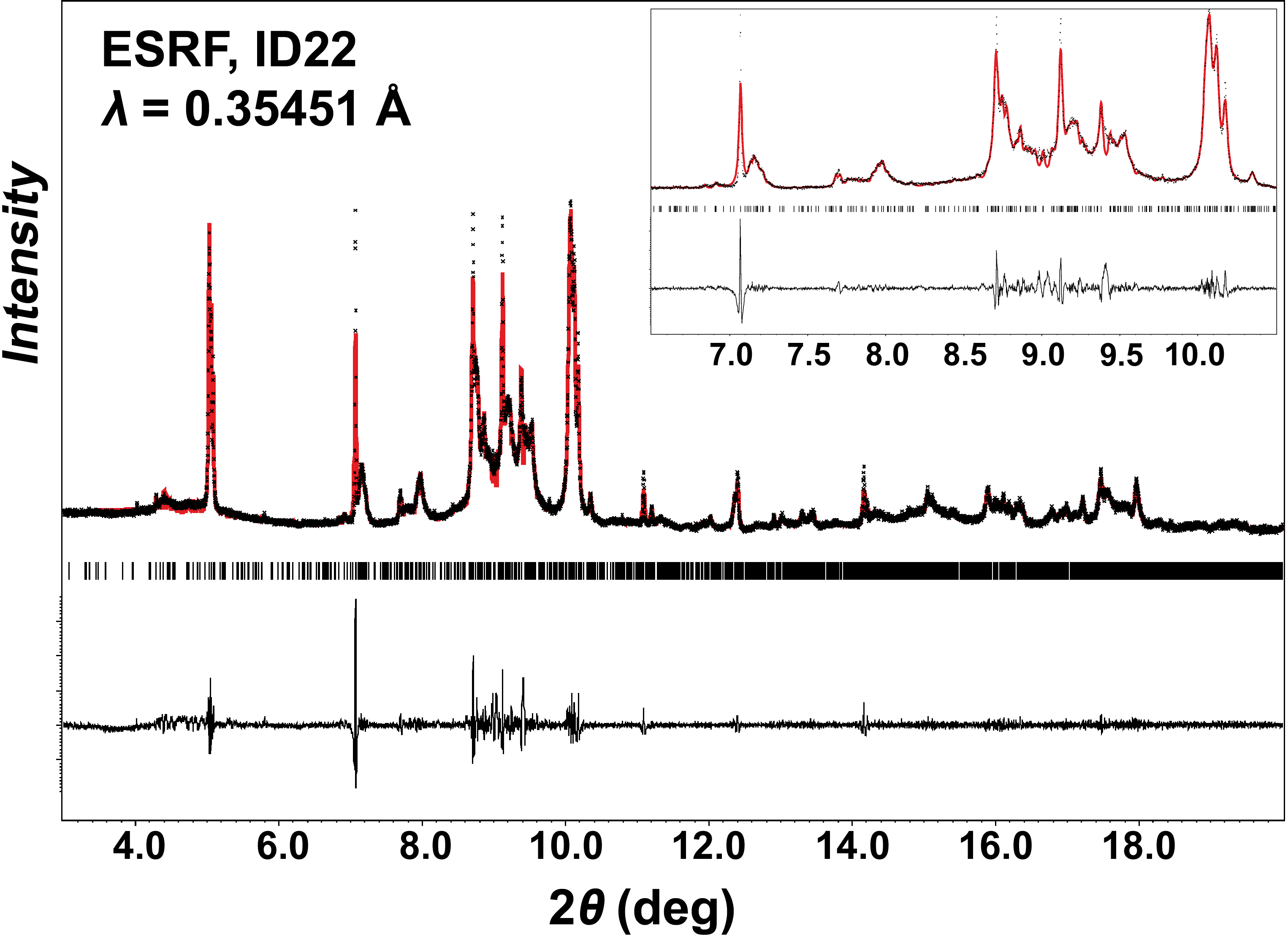}
\caption{\label{f2}Experimental (black points) and calculated (red line) HRPXRD patterns of Mo$_6$Ga$_{31}$ at room temperature. Positions of peaks are given by black ticks, and the difference plot is shown by the black line in the bottom part of figure.}
\end{figure}

Room-temperature HRPXRD pattern of polycrystalline Mo$_6$Ga$_{31}$ is presented in Figure~\ref{f2}. The pattern shows no secondary compounds, such as Mo$_8$Ga$_{41}$, Mo$_3$Ga, elemental Mo or Ga. However, all peaks show significant broadening that could be ascribed to defects within the monoclinic phase or to symmetry lowering toward the triclinic structure. Indeed, Le Bail decomposition returned similar profile $R$ factors of $R_p=6.3$, $R_{wp}=9.2$, and $GOF=2.1$ for the monoclinic unit cell, and $R_p=6.4$, $R_{wp}=9.3$, and $GOF=2.2$ for the triclinic unit cell. Under heating up to 700\,$^{\circ}$C, the HRPXRD pattern remains qualitatively the same, while reflections shift due to the monotonous increase of the lattice parameters. Both temperature-dependent HRPXRD and complementary DSC experiments show no hints for a transformation between the triclinic and monoclinic polymorphs of Mo$_6$Ga$_{31}$ (see Supporting Information).

\begin{figure}
\includegraphics{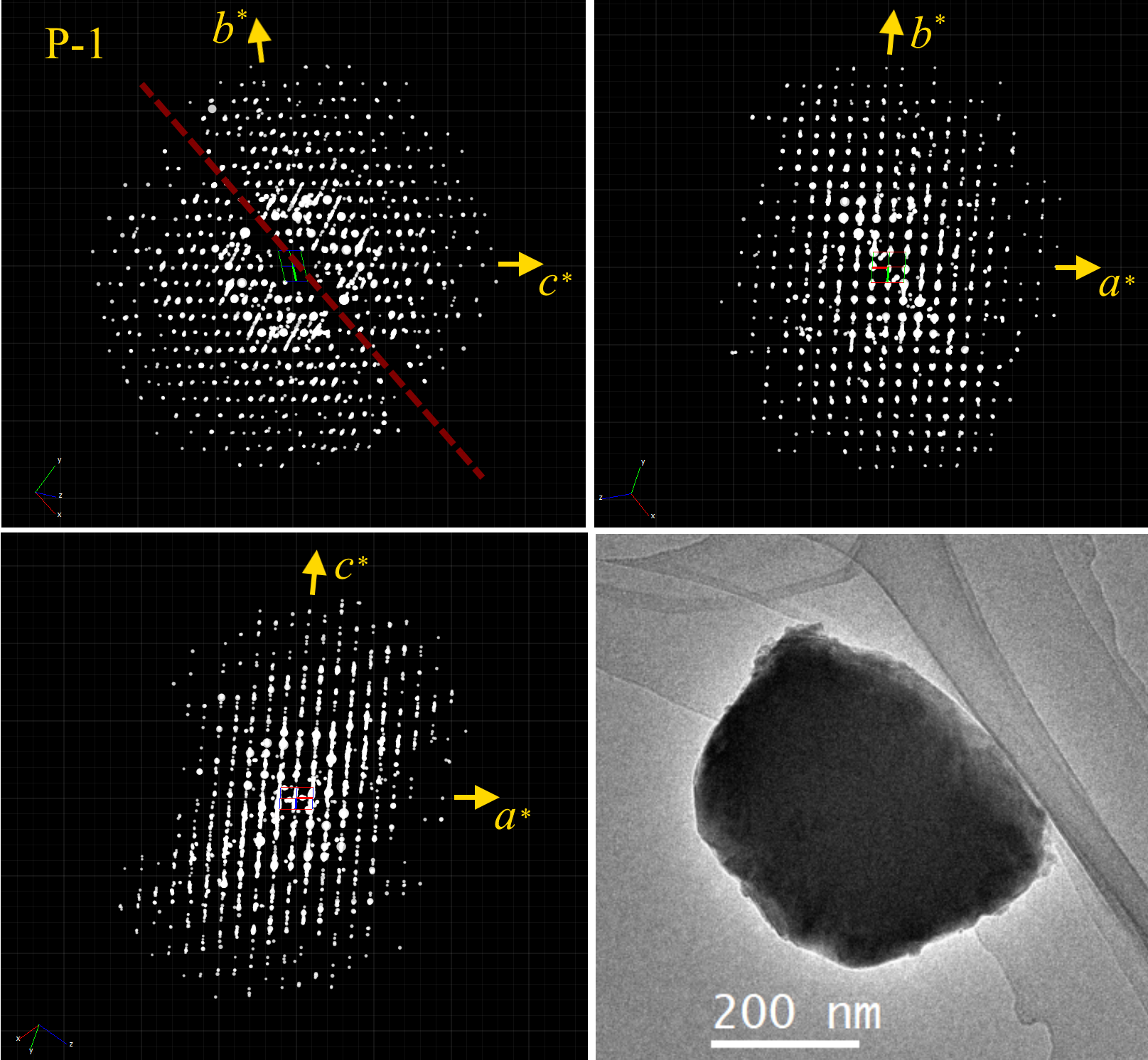}
\caption{\label{Z}Experimental 3D ED patterns of the triclinic Mo$_6$Ga$_{31}$, and the corresponding image of crystallite. Red dashed line shows the position of rotation axis towards the reciprocal space of the sample.}
\end{figure}

To gain insight into the local structure of polycrystalline Mo$_6$Ga$_{31}$, three-dimensional electron diffraction was employed providing 3D structural information from nm-size crystals. 3D ED data were collected on crystallites with the lateral size of $<500\times500$\,nm$^2$ in order to minimize the contribution of intergrowth and twinning. While testing the lattice type, two types of diffraction patterns with different unit cells were revealed corresponding to the monoclinic and triclinic modifications of Mo$_6$Ga$_{31}$. Unit cell determination by the REDp program\cite{red,xds} yields $a=9.37(4)$\,\r A, $b=16.18(6)$\,\r A, $c=16.48(1)$\,\r A, and $\beta=94.7(1)$\,$^{\circ}$ for the monoclinic Mo$_6$Ga$_{31}$, and $a=9.37(5)$\,\r A, $b=9.47(3)$\,\r A, $c=14.25(2)$\,\r A, $\alpha=85.6(2)$\,$^{\circ}$, $\beta=81.1(1)$\,$^{\circ}$, and $\gamma=85.7(1)$\,$^{\circ}$ for the triclinic Mo$_6$Ga$_{31}$. Typical 3D ED patterns of the triclinic structure are shown in Figure~\ref{Z}. The unit cell parameters are in good agreement with the single-crystal XRD results within the accuracy of the 3D ED method. EDX spectroscopy does not reveal any difference in the chemical composition of the studied crystallites (see Supporting Information). Based on the HRPXRD, DSC, and 3D ED studies, we conclude that both modifications of Mo$_6$Ga$_{31}$ are inherently present in the bulk sample.

\subsection{Electronic structure}

\begin{figure}
\includegraphics{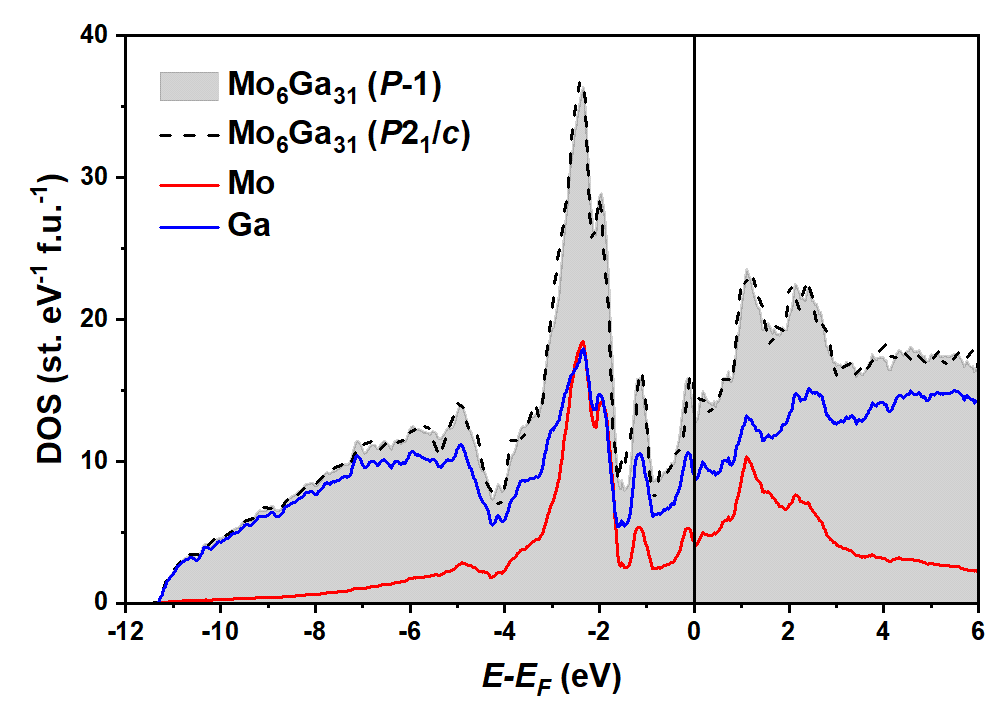}
\caption{\label{dos}Electronic structure of the triclinic Mo$_6$Ga$_{31}$. Total density of states is shown by the shaded area, the Mo and Ga contributions -- by the solid red and blue lines, respectively. The density of states of the monoclinic structure is shown by the dashed line. The position of the Fermi level is indicated by the solid black line.}
\end{figure}

Using the structural parameters obtained from the single-crystal XRD data, we calculated electronic structure of Mo$_6$Ga$_{31}$. The triclinic and monoclinic modifications possess qualitatively the same density of states (DOS), which is shown in Figure~\ref{dos}. At low energies between -12\,eV and -4\,eV, mixing of the Ga $4s$ and $4p$ states is observed with a small contribution of the Mo $4d$ states. At higher energies between -4\,eV and 4\,eV, the Ga $4p$ and Mo $4d$ states contribute to the total DOS forming the peak substructure. The Fermi level is located at the peak yielding a high value of DOS at $E=E_F$. The calculated electronic structure of Mo$_6$Ga$_{31}$ is similar to that of the Mo$_8$Ga$_{41}$,\cite{mo8ga41-2,mo8ga41-5} Mo$_4$Ga$_{21-x-\delta}$Sn$_x$,\cite{mogasn} and Mo$_7$Ga$_{52-x}$Zn$_x$\cite{mogazn} endohedral cluster compounds, which also possess high DOS at $E_F$. For Mo$_6$Ga$_{31}$, we find $N(E_F)=13$\,st. eV$^{-1}$ f.u.$^{-1}$ for the triclinic structure and 17\,st. eV$^{-1}$ f.u.$^{-1}$ for the monoclinic one. These values yield the bare Sommerfield coefficient of 31\,mJ mol$^{-1}$ K$^{-2}$ and 39\,mJ mol$^{-1}$ K$^{-2}$, respectively. For the reported endohedral cluster superconductors, the value of the density of states at the Fermi level correlates with the observed critical temperature.\cite{rega5,mogasn} Given the high DOS at $E=E_F$ calculated for Mo$_6$Ga$_{31}$, which is comparable with those of the Mo$_8$Ga$_{41}$ and Mo$_4$Ga$_{21-x-\delta}$Sn$_x$ superconductors, we expect superconducting behavior for both crystallographic modifications of Mo$_6$Ga$_{31}$. However, triclinic and monoclinic Mo$_6$Ga$_{31}$ should possess slightly different superconducting-state parameters, including the critical temperature, $T_c$, and full superconducting gap, $2\Delta(0)/(k_BT_c)$. Our HRPXRD and 3D ED studies show that the triclinic and monoclinic structures are in strong contact with each other in the bulk specimen. Due to the proximity effect,\cite{proxy} the superconducting carriers travel coherently between two superconducting phases, and thus, intermediate superconducting parameters should be observed, corresponding effectively to one superconducting phase.

\subsection{Physical properties}

Electrical resistivity of Mo$_6$Ga$_{31}$ follows metallic behavior at elevated temperatures. However, the $\rho(T)$ dependence is extremely flat between 10\,K and 300\,K, and shows the saturation behavior with increasing temperature. The small residual-resistance-ratio of 1.2 can be paralleled to the peak broadening observed in HRPXRD and caused by the coexisting domains of the monoclinic and triclinic phases. At low temperatures, superconducting transition is observed with the low-temperature drop of resistivity occuring between the onset temperature of 8.2\,K and the final temperature of 7.8\,K in zero magnetic field. The increase of magnetic field shifts the transition to lower temperatures, and finally, no indications of superconductivity are observed at temperatures above 1.8\,K in the magnetic field of $\mu_0H=10$\,T.

\begin{figure}
\includegraphics{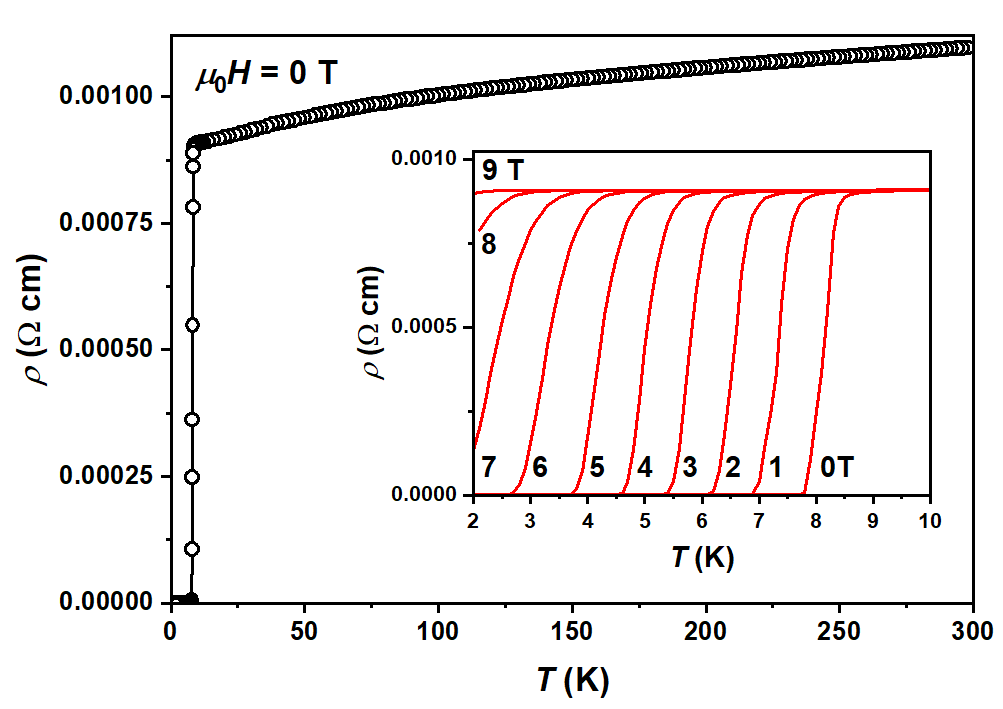}
\caption{\label{f4}Electrical resistivity of Mo$_6$Ga$_{31}$ in zero magnetic field. The inset shows the data at low temperatures in magnetic fields between 0\,T and 9\,T.}
\end{figure}

\begin{figure}
\includegraphics{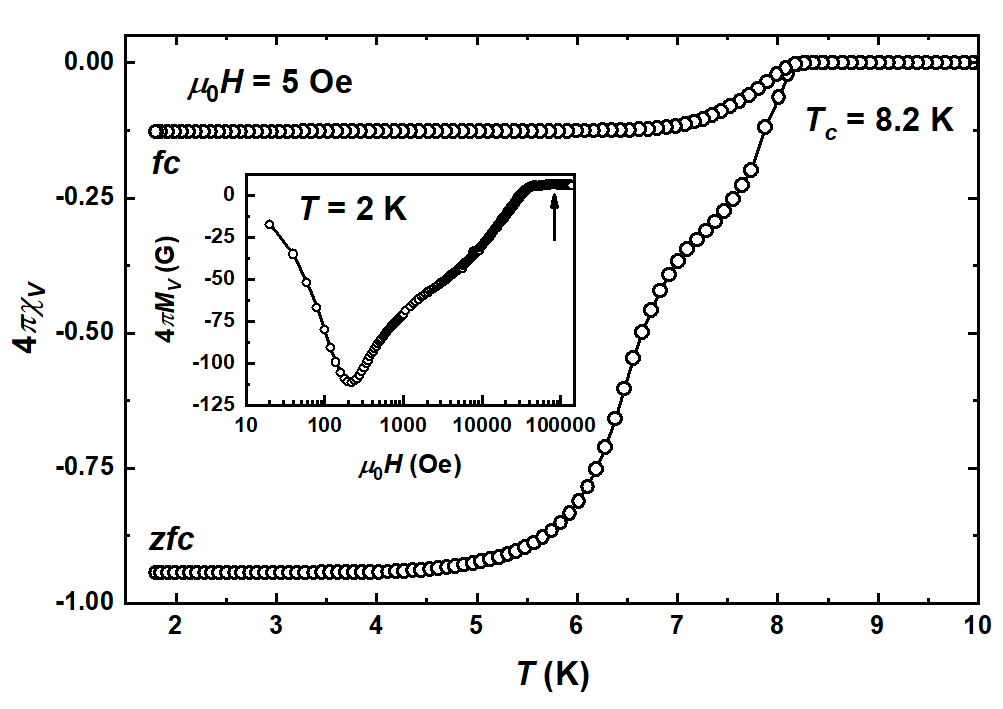}
\caption{\label{chi}Dimensionless volume magnetic susceptibility of Mo$_6$Ga$_{31}$ measured in 5\,Oe magnetic field in the zfc and fc conditions. The inset shows volume magnetization measured at $T=2$\,K. The arrow indicates the transition to the normal state.}
\end{figure}

The bulk nature of superconductivity is confirmed by thermodynamic measurements (Fig.~\ref{chi}). Dimensionless volume magnetic susceptibility shows diamagnetic shift due to the Meissner effect below the critical temperature of $T_c=8.2$\,K in 5\,Oe magnetic field. The zfc signal of -0.94 at the lowest measured temperature indicates large volume fraction of the superconducting phase. The transition is broad with temperature and shows a pronounced shoulder, which is presumably caused by the inhomogeneities of the specimen. Volume magnetization, which is shown in the inset of Figure~\ref{chi}, is characteristic of type-II superconductors. Note that the logarithmic scale is used in figure to represent the magnetic fields. In low magnetic fields, $4\pi M_V$ follows the linear behavior versus $\mu_0H$ below the lower critical field of $\mu_0H_{c1}=70$\,Oe at $T=2$\,K. With the increase of magnetic field, the normal state is achieved above the upper critical field of $\mu_0H_{c2}=6.5$\,T at $T=2$\,K. Between $\mu_0H_{c1}$ and $\mu_0H_{c2}$, the volume magnetization exhibits intricate nonmonotonic behavior, which also evidences the presence of inhomogeneities in the specimen.

\begin{figure}
\includegraphics{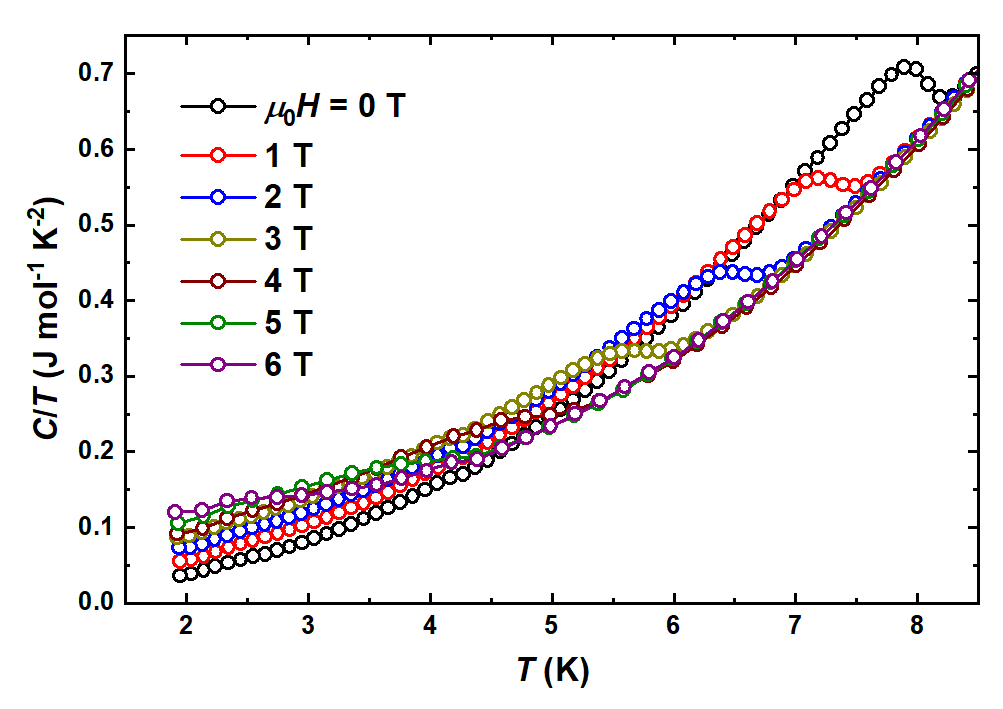}
\caption{\label{hcfields}Specific heat of Mo$_6$Ga$_{31}$ in various magnetic fields.}
\end{figure}

\begin{figure}
\includegraphics{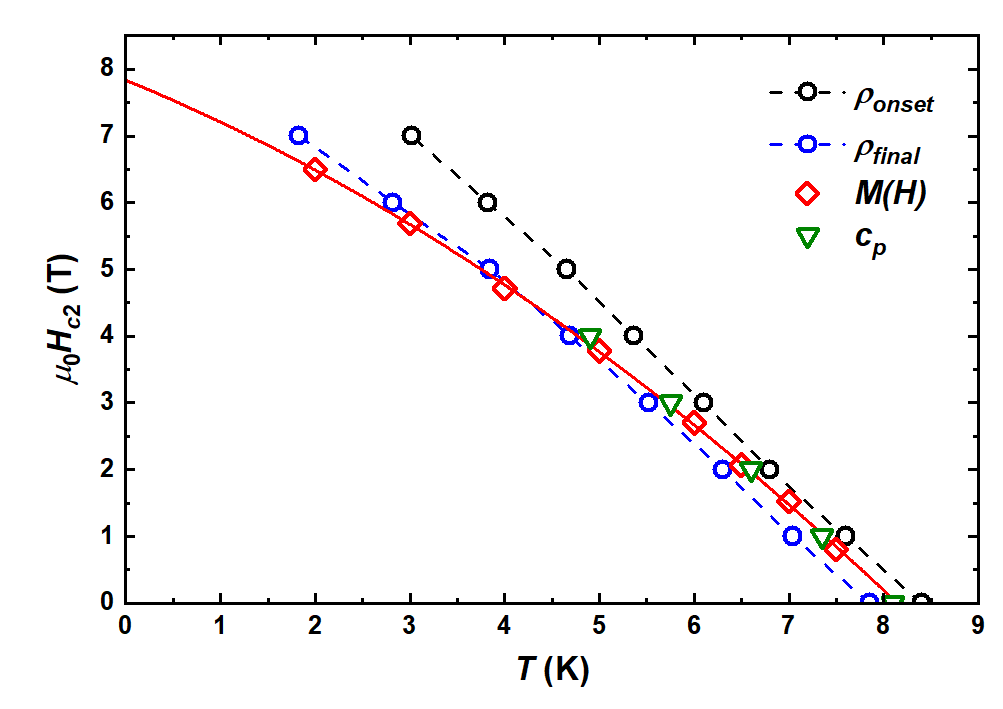}
\caption{\label{Hc2}Upper critical field of Mo$_6$Ga$_{31}$.}
\end{figure}

The specific heat of Mo$_6$Ga$_{31}$ exhibits the superconducting anomaly located at $T_c=8.2$\,K in zero magnetic field in good agreement with the resistivity and magnetization measurements. The increase of the magnetic field shifts the transition to lower temperatures, which simultaneously becomes smoothed (Fig.~\ref{hcfields}). The temperature sweeps of resistivity and heat capacity measured in various magnetic fields, as well as the field sweeps of magnetization at constant temperatures were used to extract the upper critical field of Mo$_6$Ga$_{31}$ as a function of temperature (Fig.~\ref{Hc2}). The $\mu_0H_{c2}(T)$ values, which correspond to the onset temperature of the resistivity drop, are larger than those from the magnetization and heat capacity measurements indicating the so-called $\mu_0H_{c3}(T)$ upper critical field of the surface superconductivity, which was also observed in the case of Mo$_8$Ga$_{41}$.\cite{mo8ga41-2} From the other hand, the temperatures, at which zero resistance is achieved, yields $\mu_0H_{c2}(T)$, which are closer to the thermodynamic measurements. The magnetization and heat-capacity derived $\mu_0H_{c2}(T)$ values are in good agreement with each other. Above these values, the bulk of the sample is in the normal state. Interpolation of the $\mu_0H_{c2}(T)$ values by the second-order polynomial yields $\mu_0H_{c2}(0)=7.8$\,T at zero temperature, which is in good agreement with the previous report.\cite{mo6ga31-1} The $\mu_0H_{c2}(0)$ value corresponds to the Ginzburg-Landau coherence length of $\xi=153$\,\r A as calculated from the equation $\mu_0H_{c2}(0)=\frac{\Phi_0}{2\pi\xi^2_{GL}}$, where $\Phi_0=h/2e$ is the flux quantum. At temperatures above 6\,K, $\mu_0H_{c2}(T)$ is linear with the slope of $\omega=-1.2$\,T K$^{-1}$. According to the Werthamer-Helfand-Honenberg (WHH) model, the upper critical field can be calculated as $\mu_0H_{c2}(0)=-0.693\omega{}T_{c}=6.8$\,T, which is smaller than the extrapolated value of $\mu_0H_{c2}(0)=7.8$\,T. This enhancement of the upper critical field may be due to strong electron-phonon coupling.

\begin{figure}
\includegraphics{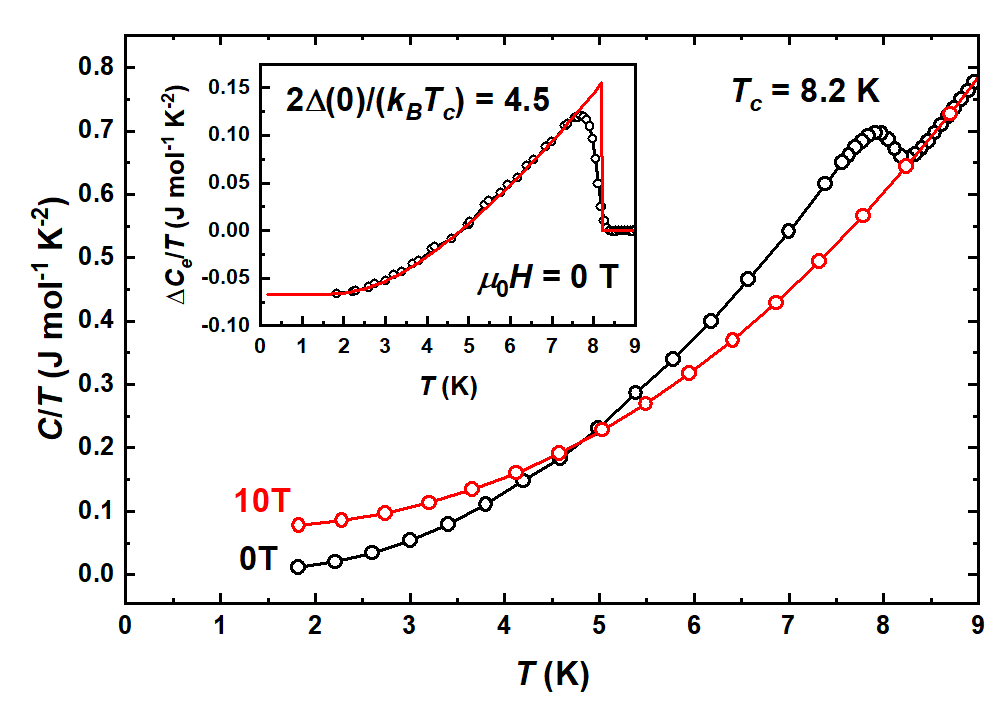}
\caption{\label{hc}Specific heat of Mo$_6$Ga$_{31}$ measured in 0\,T and 10\,T magnetic fields. The inset shows the electronic contribution to the total heat capacity in zero magnetic field. The red line is a fit according to the $\alpha$-model.}
\end{figure}

To gain insight into the electron-phonon coupling in the superconducting state of Mo$_6$Ga$_{31}$, the specific heat data were analyzed. Figure~\ref{hc} shows the specific heat in superconducting ($\mu_0H=0$\,T) and normal states ($\mu_0H=10$\,T). The electronic specific heat, which is shown in the inset of Figure~\ref{hc}, was calculated as $\Delta{}C_e/T=\Delta{}C/T(0\text{\,T})-\Delta{}C/T(10\text{\,T})$ and analyzed within the BCS-derived $\alpha$-model.\cite{alpha-1,alpha-2} The fitting yields the full superconducting gap of $2\Delta(0)/(k_BT_c)=4.5$, the normalized specific heat jump of $\Delta{}C_e/(\gamma_NT_c)=2.3$ at $T=T_c$, and the normal-state Sommerfield coefficient of $\gamma_N=67$\,mJ mol$^{-1}$ K$^{-2}$. Remarkably, the transition is smoothed with temperature, which is seen as the difference between the experimental data (black open circles in the inset of Fig.~\ref{hc}) and the calculated specific heat (solid red line) in the vicinity of the critical temperature, which may be caused by the coexisting domains of the monoclinic and triclinic phases. Both the enhanced value of $\Delta{}C_e/(\gamma_NT_c)=2.3$, which is larger than the weak-coupling BCS limit of $\Delta{}C_e/(\gamma_NT_c)=1.43$, and the large value of $\alpha=2.25$ point to the strong electron-phonon coupling in the superconducting state. Using the bare Sommerfield coefficient of $\gamma_{bare}=31.1$\,mJ mol$^{-1}$ K$^{-2}$ (triclinic structure) and 39.1\,mJ mol$^{-1}$ K$^{-2}$ (monoclinic structure), the electron-phonon coupling constant can be calculated as $\lambda_{ep}=\frac{\gamma_N}{\gamma_{bare}}-1$, which yields $\lambda_{ep}=1.15$ and 0.7 for the triclinic and monoclinic modifications, respectively. Both values suggest strong or moderate electron-phonon coupling. Thus, Mo$_6$Ga$_{31}$ is a strongly-coupled superconductor similar to the Mo$_8$Ga$_{41}$ and Mo$_4$Ga$_{21-x-\delta}$Sn$_x$ related compounds.\cite{mo8ga41-2,mo8ga41-3,mogasn}

\begin{table}
  \caption{\ Normal- and superconducting-state parameters of the Mo$_8$Ga$_{41}$,\cite{mo8ga41-2,mo8ga41-3} Mo$_6$Ga$_{31}$, and Mo$_4$Ga$_{21-x-\delta}$Sn$_x$\cite{mogasn} endohedral cluster superconductors. $m$ refers to the monoclinic polymorph of Mo$_6$Ga$_{31}$, and $t$ to triclinic.}
  \label{t1}
  \begin{tabular}{llll}
    \hline
    Parameter & Mo$_8$Ga$_{41}$ & Mo$_6$Ga$_{31}$ & Mo$_4$Ga$_{21-x-\delta}$Sn$_x$ \\
    \hline
    $T_c$ (K) & 9.8 & 8.2 & 5.85 \\
    VEC (e per Mo) & 21.375 & 21.5 & 21.85 \\
    $\mu_0H_{c2}$ (T) at $T=2$\,K & 7.45 & 6.5 & 1.9 \\
    $2\Delta(0)/(k_BT_c)$ & 4.4 & 4.5 & 4.1 \\
    $\gamma_N$ (mJ mol$^{-1}$ K$^{-2}$) & 99 & 67 & 39 \\
    $\gamma_{bare}$ (mJ mol$^{-1}$ K$^{-2}$) & 52.7 & 31.1$^t$/39.1$^m$ & 20.0 \\
    $\lambda_{ep}$ & 0.9 & 1.15$^t$/0.7$^m$ & 0.95 \\
    \hline
  \end{tabular}
\end{table}

In Table~\ref{t1}, we compare normal- and superconducting-state properties of Mo$_8$Ga$_{41}$, Mo$_6$Ga$_{31}$, and Mo$_4$Ga$_{21-x-\delta}$Sn$_x$. The observed $T_c$ values follow the trend of decreasing the critical temperature with increasing the valence electron count (VEC), which was proposed for endohedral cluster superconductors.\cite{rega5} Moreover, it is obvious that the critical temperature correlates with the value of DOS at $E=E_F$: both $\gamma_N$ and $\gamma_{bare}$ decrease with decreasing $T_c$. At the same time, the values of full superconducting gap, $2\Delta(0)/(k_BT_c)$, and electron-phonon coupling constant, $\lambda_{ep}$, almost do not vary in the series indicating that the strength of the electron-phonon coupling in the superconducting state has no significant impact on the critical temperature.

\section{Conclusions}

In the paper, synthesis, structural characteristics, computational electronic structure, and physical properties of the Mo$_6$Ga$_{31}$ superconductor are reported. The bulk type-II superconductivity is observed below the critical temperature of 8.2\,K in zero magnetic field, and below the upper critical field of 7.8\,T extrapolated to zero temperature. Remarkably, the superconducting state is in the strong-coupling regime with the large ratio of $2\Delta/(k_BT_c)=4.5$. This places Mo$_6$Ga$_{31}$ in line with the Mo$_8$Ga$_{41}$ and  Mo$_4$Ga$_{21-x-\delta}$Sn$_x$ superconductors, for which the strong electron-phonon coupling was also reported. Interestingly, the critical temperture correlates with the composition in the Mo$_8$Ga$_{41}$ ($T_c=9.8$\,K), Mo$_6$Ga$_{31}$ (8.2\,K), and Mo$_4$Ga$_{21-x-\delta}$Sn$_x$ (5.85\,K) series, namely, with the number of Mo atoms per formula unit, and with the density of states at the Fermi level. In the listed superconductors, Mo is placed inside the Mo@Ga$_{10}$ endohedral clusters, where strong mixing of the Mo $4d$ states and Ga $4p$ states yields peculiar electronic structure in the vicinity of the Fermi energy. The high density of states is observed at $E=E_F$, which may be favorable for superconductivty. Moreover, the framework composed of Mo@Ga$_{10}$ clusters exhibits specific phonon properties that give rise to the strong electron-phonon coupling. Thus, Mo$_6$Ga$_{31}$ expands the family of Mo-based strongly-coupled superconductors.

\begin{acknowledgements}
The authors acknowledge the European Synchrotron Radiation Facility for granting the beam time and thank Dr. Wilson Mogodi for his help during the high-resolution PXRD experiments. The work was supported by the Russian Science Foundation, grant no. 17-13-01033. V.Yu.V. acknowledges the financial support from the Mobilitas program of the European Science Foundation, grant no. MOBJD449. A.A.T. appreciates financial support by the Federal Ministry for Education and Research under the Sofja Kovalevskaya Award of the Alexander von Humboldt Foundation. Z.W. and E.V.D. thank the National Science Foundation for supporting structural studies under grant no. CHE-1152441.
\end{acknowledgements}

\bibliography{fulltext}

\begin{thebibliography}{26}%
\makeatletter
\providecommand \@ifxundefined [1]{%
 \@ifx{#1\undefined}
}%
\providecommand \@ifnum [1]{%
 \ifnum #1\expandafter \@firstoftwo
 \else \expandafter \@secondoftwo
 \fi
}%
\providecommand \@ifx [1]{%
 \ifx #1\expandafter \@firstoftwo
 \else \expandafter \@secondoftwo
 \fi
}%
\providecommand \natexlab [1]{#1}%
\providecommand \enquote  [1]{``#1''}%
\providecommand \bibnamefont  [1]{#1}%
\providecommand \bibfnamefont [1]{#1}%
\providecommand \citenamefont [1]{#1}%
\providecommand \href@noop [0]{\@secondoftwo}%
\providecommand \href [0]{\begingroup \@sanitize@url \@href}%
\providecommand \@href[1]{\@@startlink{#1}\@@href}%
\providecommand \@@href[1]{\endgroup#1\@@endlink}%
\providecommand \@sanitize@url [0]{\catcode `\\12\catcode `\$12\catcode
  `\&12\catcode `\#12\catcode `\^12\catcode `\_12\catcode `\%12\relax}%
\providecommand \@@startlink[1]{}%
\providecommand \@@endlink[0]{}%
\providecommand \url  [0]{\begingroup\@sanitize@url \@url }%
\providecommand \@url [1]{\endgroup\@href {#1}{\urlprefix }}%
\providecommand \urlprefix  [0]{URL }%
\providecommand \Eprint [0]{\href }%
\providecommand \doibase [0]{http://dx.doi.org/}%
\providecommand \selectlanguage [0]{\@gobble}%
\providecommand \bibinfo  [0]{\@secondoftwo}%
\providecommand \bibfield  [0]{\@secondoftwo}%
\providecommand \translation [1]{[#1]}%
\providecommand \BibitemOpen [0]{}%
\providecommand \bibitemStop [0]{}%
\providecommand \bibitemNoStop [0]{.\EOS\space}%
\providecommand \EOS [0]{\spacefactor3000\relax}%
\providecommand \BibitemShut  [1]{\csname bibitem#1\endcsname}%
\let\auto@bib@innerbib\@empty
\bibitem [{\citenamefont {Xie}\ \emph {et~al.}(2015)\citenamefont {Xie},
  \citenamefont {Luo}, \citenamefont {Phelan}, \citenamefont {Klimczuk},
  \citenamefont {Cevallos},\ and\ \citenamefont {Cava}}]{rega5}%
  \BibitemOpen
  \bibfield  {author} {\bibinfo {author} {\bibfnamefont {W.}~\bibnamefont
  {Xie}}, \bibinfo {author} {\bibfnamefont {H.}~\bibnamefont {Luo}}, \bibinfo
  {author} {\bibfnamefont {B.~F.}\ \bibnamefont {Phelan}}, \bibinfo {author}
  {\bibfnamefont {T.}~\bibnamefont {Klimczuk}}, \bibinfo {author}
  {\bibfnamefont {F.~A.}\ \bibnamefont {Cevallos}}, \ and\ \bibinfo {author}
  {\bibfnamefont {R.~J.}\ \bibnamefont {Cava}},\ }\href {\doibase
  10.1073/pnas.1522191112} {\bibfield  {journal} {\bibinfo  {journal} {{P}roc.
  {N}atl. {A}cad. {S}ci.{ USA}}\ }\textbf {\bibinfo {volume} {112}},\ \bibinfo
  {pages} {E7048} (\bibinfo {year} {2015})}\BibitemShut {NoStop}%
\bibitem [{\citenamefont {Bezinge}\ \emph {et~al.}(1984)\citenamefont
  {Bezinge}, \citenamefont {Yvon}, \citenamefont {Decroux},\ and\ \citenamefont
  {Muller}}]{mo8ga41-1}%
  \BibitemOpen
  \bibfield  {author} {\bibinfo {author} {\bibfnamefont {A.}~\bibnamefont
  {Bezinge}}, \bibinfo {author} {\bibfnamefont {K.}~\bibnamefont {Yvon}},
  \bibinfo {author} {\bibfnamefont {M.}~\bibnamefont {Decroux}}, \ and\
  \bibinfo {author} {\bibfnamefont {J.}~\bibnamefont {Muller}},\ }\href
  {\doibase 10.1016/0022-5088(84)90237-6} {\bibfield  {journal} {\bibinfo
  {journal} {{J}. {L}ess-{C}ommon {M}et.}\ }\textbf {\bibinfo {volume} {99}},\
  \bibinfo {pages} {L27} (\bibinfo {year} {1984})}\BibitemShut {NoStop}%
\bibitem [{\citenamefont {Fischer}(1972)}]{mo6ga31-1}%
  \BibitemOpen
  \bibfield  {author} {\bibinfo {author} {\bibfnamefont {O.}~\bibnamefont
  {Fischer}},\ }\href {\doibase 10.5169/seals-114388} {\bibfield  {journal}
  {\bibinfo  {journal} {{H}elvetica {P}hysica {A}cta}\ }\textbf {\bibinfo
  {volume} {45}},\ \bibinfo {pages} {331} (\bibinfo {year} {1972})}\BibitemShut
  {NoStop}%
\bibitem [{\citenamefont {Verchenko}\ \emph
  {et~al.}(2019{\natexlab{a}})\citenamefont {Verchenko}, \citenamefont
  {Zubtsovskii}, \citenamefont {Wei}, \citenamefont {Tsirlin}, \citenamefont
  {Marcin}, \citenamefont {Sobolev}, \citenamefont {Presniakov}, \citenamefont
  {Dikarev},\ and\ \citenamefont {Shevelkov}}]{mogasn}%
  \BibitemOpen
  \bibfield  {author} {\bibinfo {author} {\bibfnamefont {V.~Y.}\ \bibnamefont
  {Verchenko}}, \bibinfo {author} {\bibfnamefont {A.~O.}\ \bibnamefont
  {Zubtsovskii}}, \bibinfo {author} {\bibfnamefont {Z.}~\bibnamefont {Wei}},
  \bibinfo {author} {\bibfnamefont {A.~A.}\ \bibnamefont {Tsirlin}}, \bibinfo
  {author} {\bibfnamefont {M.}~\bibnamefont {Marcin}}, \bibinfo {author}
  {\bibfnamefont {A.~V.}\ \bibnamefont {Sobolev}}, \bibinfo {author}
  {\bibfnamefont {I.~A.}\ \bibnamefont {Presniakov}}, \bibinfo {author}
  {\bibfnamefont {E.~V.}\ \bibnamefont {Dikarev}}, \ and\ \bibinfo {author}
  {\bibfnamefont {A.~V.}\ \bibnamefont {Shevelkov}},\ }\href {\doibase
  10.1021/acs.inorgchem.9b02598} {\bibfield  {journal} {\bibinfo  {journal}
  {{I}norg. {C}hem.}\ }\textbf {\bibinfo {volume} {58}},\ \bibinfo {pages}
  {15552} (\bibinfo {year} {2019}{\natexlab{a}})}\BibitemShut {NoStop}%
\bibitem [{\citenamefont {Shibayama}\ \emph {et~al.}(2007)\citenamefont
  {Shibayama}, \citenamefont {Nohara}, \citenamefont {Katori}, \citenamefont
  {Okamoto}, \citenamefont {Hiroi},\ and\ \citenamefont {Takagi}}]{t2ga9}%
  \BibitemOpen
  \bibfield  {author} {\bibinfo {author} {\bibfnamefont {T.}~\bibnamefont
  {Shibayama}}, \bibinfo {author} {\bibfnamefont {M.}~\bibnamefont {Nohara}},
  \bibinfo {author} {\bibfnamefont {H.~A.}\ \bibnamefont {Katori}}, \bibinfo
  {author} {\bibfnamefont {Y.}~\bibnamefont {Okamoto}}, \bibinfo {author}
  {\bibfnamefont {Z.}~\bibnamefont {Hiroi}}, \ and\ \bibinfo {author}
  {\bibfnamefont {H.}~\bibnamefont {Takagi}},\ }\href {\doibase
  10.1143/JPSJ.76.073708} {\bibfield  {journal} {\bibinfo  {journal} {{J}.
  {P}hys. {S}oc. {J}pn.}\ }\textbf {\bibinfo {volume} {76}},\ \bibinfo {pages}
  {073708} (\bibinfo {year} {2007})}\BibitemShut {NoStop}%
\bibitem [{\citenamefont {Verchenko}\ \emph {et~al.}(2016)\citenamefont
  {Verchenko}, \citenamefont {Tsirlin}, \citenamefont {Zubtsovskiy},\ and\
  \citenamefont {Shevelkov}}]{mo8ga41-2}%
  \BibitemOpen
  \bibfield  {author} {\bibinfo {author} {\bibfnamefont {V.~Y.}\ \bibnamefont
  {Verchenko}}, \bibinfo {author} {\bibfnamefont {A.~A.}\ \bibnamefont
  {Tsirlin}}, \bibinfo {author} {\bibfnamefont {A.~O.}\ \bibnamefont
  {Zubtsovskiy}}, \ and\ \bibinfo {author} {\bibfnamefont {A.~V.}\ \bibnamefont
  {Shevelkov}},\ }\href {\doibase 10.1103/PhysRevB.93.064501} {\bibfield
  {journal} {\bibinfo  {journal} {{P}hys. {R}ev. {B}}\ }\textbf {\bibinfo
  {volume} {93}},\ \bibinfo {pages} {064501} (\bibinfo {year}
  {2016})}\BibitemShut {NoStop}%
\bibitem [{\citenamefont {Marcin}\ \emph {et~al.}(2019)\citenamefont {Marcin},
  \citenamefont {Ka\v{c}mar\v{c}\'{i}k}, \citenamefont {Pribulov\'{a}},
  \citenamefont {Kop\v{c}\'{i}k}, \citenamefont {Szab\'{o}}, \citenamefont
  {\v{S}ofranko}, \citenamefont {Samuely}, \citenamefont {Va\v{n}o},
  \citenamefont {Marcenat}, \citenamefont {Verchenko}, \citenamefont
  {Shevelkov},\ and\ \citenamefont {Samuely}}]{mo8ga41-3}%
  \BibitemOpen
  \bibfield  {author} {\bibinfo {author} {\bibfnamefont {M.}~\bibnamefont
  {Marcin}}, \bibinfo {author} {\bibfnamefont {J.}~\bibnamefont
  {Ka\v{c}mar\v{c}\'{i}k}}, \bibinfo {author} {\bibfnamefont {Z.}~\bibnamefont
  {Pribulov\'{a}}}, \bibinfo {author} {\bibfnamefont {M.}~\bibnamefont
  {Kop\v{c}\'{i}k}}, \bibinfo {author} {\bibfnamefont {P.}~\bibnamefont
  {Szab\'{o}}}, \bibinfo {author} {\bibfnamefont {O.}~\bibnamefont
  {\v{S}ofranko}}, \bibinfo {author} {\bibfnamefont {T.}~\bibnamefont
  {Samuely}}, \bibinfo {author} {\bibfnamefont {V.}~\bibnamefont {Va\v{n}o}},
  \bibinfo {author} {\bibfnamefont {C.}~\bibnamefont {Marcenat}}, \bibinfo
  {author} {\bibfnamefont {V.~Y.}\ \bibnamefont {Verchenko}}, \bibinfo {author}
  {\bibfnamefont {A.~V.}\ \bibnamefont {Shevelkov}}, \ and\ \bibinfo {author}
  {\bibfnamefont {P.}~\bibnamefont {Samuely}},\ }\href {\doibase
  10.1038/s41598-019-49846-y} {\bibfield  {journal} {\bibinfo  {journal}
  {{S}ci. {R}ep.}\ }\textbf {\bibinfo {volume} {9}},\ \bibinfo {pages} {13552}
  (\bibinfo {year} {2019})}\BibitemShut {NoStop}%
\bibitem [{\citenamefont {Verchenko}\ \emph {et~al.}(2017)\citenamefont
  {Verchenko}, \citenamefont {Khasanov}, \citenamefont {Guguchia},
  \citenamefont {Tsirlin},\ and\ \citenamefont {Shevelkov}}]{mo8ga41-4}%
  \BibitemOpen
  \bibfield  {author} {\bibinfo {author} {\bibfnamefont {V.~Y.}\ \bibnamefont
  {Verchenko}}, \bibinfo {author} {\bibfnamefont {R.}~\bibnamefont {Khasanov}},
  \bibinfo {author} {\bibfnamefont {Z.}~\bibnamefont {Guguchia}}, \bibinfo
  {author} {\bibfnamefont {A.~A.}\ \bibnamefont {Tsirlin}}, \ and\ \bibinfo
  {author} {\bibfnamefont {A.~V.}\ \bibnamefont {Shevelkov}},\ }\href {\doibase
  10.1103/PhysRevB.96.134504} {\bibfield  {journal} {\bibinfo  {journal}
  {{P}hys. {R}ev. {B}}\ }\textbf {\bibinfo {volume} {96}},\ \bibinfo {pages}
  {134504} (\bibinfo {year} {2017})}\BibitemShut {NoStop}%
\bibitem [{\citenamefont {Sirohi}\ \emph {et~al.}(2019)\citenamefont {Sirohi},
  \citenamefont {Saha}, \citenamefont {Neha}, \citenamefont {Das},
  \citenamefont {Patnaik}, \citenamefont {Das},\ and\ \citenamefont
  {Sheet}}]{mo8ga41-5}%
  \BibitemOpen
  \bibfield  {author} {\bibinfo {author} {\bibfnamefont {A.}~\bibnamefont
  {Sirohi}}, \bibinfo {author} {\bibfnamefont {S.}~\bibnamefont {Saha}},
  \bibinfo {author} {\bibfnamefont {P.}~\bibnamefont {Neha}}, \bibinfo {author}
  {\bibfnamefont {S.}~\bibnamefont {Das}}, \bibinfo {author} {\bibfnamefont
  {S.}~\bibnamefont {Patnaik}}, \bibinfo {author} {\bibfnamefont
  {T.}~\bibnamefont {Das}}, \ and\ \bibinfo {author} {\bibfnamefont
  {G.}~\bibnamefont {Sheet}},\ }\href {\doibase 10.1103/PhysRevB.99.054503}
  {\bibfield  {journal} {\bibinfo  {journal} {{P}hys. {R}ev. {B}}\ }\textbf
  {\bibinfo {volume} {99}},\ \bibinfo {pages} {054503} (\bibinfo {year}
  {2019})}\BibitemShut {NoStop}%
\bibitem [{\citenamefont {Yvon}(1974)}]{mo6ga31-2}%
  \BibitemOpen
  \bibfield  {author} {\bibinfo {author} {\bibfnamefont {K.}~\bibnamefont
  {Yvon}},\ }\href {\doibase 10.1107/S0567740874010958} {\bibfield  {journal}
  {\bibinfo  {journal} {{A}cta {C}ryst. {B}}\ }\textbf {\bibinfo {volume}
  {30}},\ \bibinfo {pages} {853} (\bibinfo {year} {1974})}\BibitemShut
  {NoStop}%
\bibitem [{\citenamefont {Lux}(2004)}]{mo6ga31-3}%
  \BibitemOpen
  \bibfield  {author} {\bibinfo {author} {\bibfnamefont {R.}~\bibnamefont
  {Lux}},\ }\href@noop {} {\emph {\bibinfo {title} {{D}issertation:
  {I}ntermetallische {V}erbindungen mit hochschmelzenden
  {\"{U}}bergangsmetallen und niedrigschmelzenden {M}etallen}}}\ (\bibinfo
  {publisher} {University of Freiburg},\ \bibinfo {address} {Freiburg,
  Germany},\ \bibinfo {year} {2004})\BibitemShut {NoStop}%
\bibitem [{\citenamefont {Pet\v{r}\'{i}\v{c}ek}\ \emph
  {et~al.}(2014)\citenamefont {Pet\v{r}\'{i}\v{c}ek}, \citenamefont
  {Du\v{s}ek},\ and\ \citenamefont {Palatinus}}]{jana}%
  \BibitemOpen
  \bibfield  {author} {\bibinfo {author} {\bibfnamefont {V.}~\bibnamefont
  {Pet\v{r}\'{i}\v{c}ek}}, \bibinfo {author} {\bibfnamefont {M.}~\bibnamefont
  {Du\v{s}ek}}, \ and\ \bibinfo {author} {\bibfnamefont {L.}~\bibnamefont
  {Palatinus}},\ }\href {\doibase 10.1515/zkri-2014-1737} {\bibfield  {journal}
  {\bibinfo  {journal} {{Z}. {K}ristallogr.}\ }\textbf {\bibinfo {volume}
  {229}},\ \bibinfo {pages} {345} (\bibinfo {year} {2014})}\BibitemShut
  {NoStop}%
\bibitem [{\citenamefont {Roslova}\ \emph {et~al.}(2019)\citenamefont
  {Roslova}, \citenamefont {Smeets}, \citenamefont {Wang}, \citenamefont
  {Thersleff}, \citenamefont {Xu},\ and\ \citenamefont {Zou}}]{instead}%
  \BibitemOpen
  \bibfield  {author} {\bibinfo {author} {\bibfnamefont {M.}~\bibnamefont
  {Roslova}}, \bibinfo {author} {\bibfnamefont {S.}~\bibnamefont {Smeets}},
  \bibinfo {author} {\bibfnamefont {B.}~\bibnamefont {Wang}}, \bibinfo {author}
  {\bibfnamefont {T.}~\bibnamefont {Thersleff}}, \bibinfo {author}
  {\bibfnamefont {H.}~\bibnamefont {Xu}}, \ and\ \bibinfo {author}
  {\bibfnamefont {X.}~\bibnamefont {Zou}},\ }\href@noop {} {\enquote {\bibinfo
  {title} {Towards cross-platform automated rotation electron diffraction},}\ }
  (\bibinfo {year} {2019}),\ \Eprint {http://arxiv.org/abs/arXiv:1911.09393}
  {arXiv:1911.09393} \BibitemShut {NoStop}%
\bibitem [{\citenamefont {Wan}\ \emph {et~al.}(2013)\citenamefont {Wan},
  \citenamefont {Sun}, \citenamefont {Su}, \citenamefont {Hovm\"{o}ller},\ and\
  \citenamefont {Zou}}]{red}%
  \BibitemOpen
  \bibfield  {author} {\bibinfo {author} {\bibfnamefont {W.}~\bibnamefont
  {Wan}}, \bibinfo {author} {\bibfnamefont {J.}~\bibnamefont {Sun}}, \bibinfo
  {author} {\bibfnamefont {J.}~\bibnamefont {Su}}, \bibinfo {author}
  {\bibfnamefont {S.}~\bibnamefont {Hovm\"{o}ller}}, \ and\ \bibinfo {author}
  {\bibfnamefont {X.}~\bibnamefont {Zou}},\ }\href {\doibase
  10.1107/S0021889813027714} {\bibfield  {journal} {\bibinfo  {journal} {{J}.
  {A}ppl. {C}ryst.}\ }\textbf {\bibinfo {volume} {46}},\ \bibinfo {pages}
  {1863} (\bibinfo {year} {2013})}\BibitemShut {NoStop}%
\bibitem [{\citenamefont {Palatinus}\ and\ \citenamefont
  {Chapius}(2007)}]{superflip}%
  \BibitemOpen
  \bibfield  {author} {\bibinfo {author} {\bibfnamefont {L.}~\bibnamefont
  {Palatinus}}\ and\ \bibinfo {author} {\bibfnamefont {G.}~\bibnamefont
  {Chapius}},\ }\href {\doibase 10.1107/S0021889807029238} {\bibfield
  {journal} {\bibinfo  {journal} {{J}. {A}ppl. {C}ryst.}\ }\textbf {\bibinfo
  {volume} {40}},\ \bibinfo {pages} {786} (\bibinfo {year} {2007})}\BibitemShut
  {NoStop}%
\bibitem [{\citenamefont {Sheldrick}(2015)}]{shelx}%
  \BibitemOpen
  \bibfield  {author} {\bibinfo {author} {\bibfnamefont {G.~M.}\ \bibnamefont
  {Sheldrick}},\ }\href {\doibase 10.1107/S2053229614024218} {\bibfield
  {journal} {\bibinfo  {journal} {{A}cta {C}rystallogr., {S}ect. {C}: {S}truct.
  {C}hem.}\ }\textbf {\bibinfo {volume} {C71}},\ \bibinfo {pages} {3} (\bibinfo
  {year} {2015})}\BibitemShut {NoStop}%
\bibitem [{\citenamefont {Gelato}\ and\ \citenamefont
  {Parth{\'e}}(1987)}]{tidy}%
  \BibitemOpen
  \bibfield  {author} {\bibinfo {author} {\bibfnamefont {L.~M.}\ \bibnamefont
  {Gelato}}\ and\ \bibinfo {author} {\bibfnamefont {E.}~\bibnamefont
  {Parth{\'e}}},\ }\href {\doibase 10.1107/S0021889887086965} {\bibfield
  {journal} {\bibinfo  {journal} {J. Appl. Cryst.}\ }\textbf {\bibinfo {volume}
  {20}},\ \bibinfo {pages} {139} (\bibinfo {year} {1987})}\BibitemShut
  {NoStop}%
\bibitem [{\citenamefont {Momma}\ and\ \citenamefont {Izumi}(2011)}]{vesta}%
  \BibitemOpen
  \bibfield  {author} {\bibinfo {author} {\bibfnamefont {K.}~\bibnamefont
  {Momma}}\ and\ \bibinfo {author} {\bibfnamefont {F.}~\bibnamefont {Izumi}},\
  }\href {\doibase 10.1107/S0021889811038970} {\bibfield  {journal} {\bibinfo
  {journal} {{J}. {A}ppl. {C}ryst.}\ }\textbf {\bibinfo {volume} {44}},\
  \bibinfo {pages} {1272} (\bibinfo {year} {2011})}\BibitemShut {NoStop}%
\bibitem [{\citenamefont {Koepernik}\ and\ \citenamefont
  {Eschrig}(1999)}]{fplo}%
  \BibitemOpen
  \bibfield  {author} {\bibinfo {author} {\bibfnamefont {K.}~\bibnamefont
  {Koepernik}}\ and\ \bibinfo {author} {\bibfnamefont {H.}~\bibnamefont
  {Eschrig}},\ }\href {\doibase 10.1103/PhysRevB.59.1743} {\bibfield  {journal}
  {\bibinfo  {journal} {{P}hys. {R}ev. {B}}\ }\textbf {\bibinfo {volume}
  {59}},\ \bibinfo {pages} {1743} (\bibinfo {year} {1999})}\BibitemShut
  {NoStop}%
\bibitem [{\citenamefont {Perdew}\ and\ \citenamefont {Wang}(1992)}]{lda}%
  \BibitemOpen
  \bibfield  {author} {\bibinfo {author} {\bibfnamefont {J.~P.}\ \bibnamefont
  {Perdew}}\ and\ \bibinfo {author} {\bibfnamefont {Y.}~\bibnamefont {Wang}},\
  }\href {\doibase 10.1103/PhysRevB.45.13244} {\bibfield  {journal} {\bibinfo
  {journal} {{P}hys. {R}ev. {B}}\ }\textbf {\bibinfo {volume} {45}},\ \bibinfo
  {pages} {13244} (\bibinfo {year} {1992})}\BibitemShut {NoStop}%
\bibitem [{\citenamefont {Bl\"{o}chl}\ \emph {et~al.}(1994)\citenamefont
  {Bl\"{o}chl}, \citenamefont {Jepsen},\ and\ \citenamefont
  {Andersen}}]{integr}%
  \BibitemOpen
  \bibfield  {author} {\bibinfo {author} {\bibfnamefont {P.~E.}\ \bibnamefont
  {Bl\"{o}chl}}, \bibinfo {author} {\bibfnamefont {O.}~\bibnamefont {Jepsen}},
  \ and\ \bibinfo {author} {\bibfnamefont {O.~K.}\ \bibnamefont {Andersen}},\
  }\href {\doibase 10.1103/PhysRevB.49.16223} {\bibfield  {journal} {\bibinfo
  {journal} {{P}hys. {R}ev. {B}}\ }\textbf {\bibinfo {volume} {49}},\ \bibinfo
  {pages} {16223} (\bibinfo {year} {1994})}\BibitemShut {NoStop}%
\bibitem [{\citenamefont {Kabsch}(2010)}]{xds}%
  \BibitemOpen
  \bibfield  {author} {\bibinfo {author} {\bibfnamefont {W.}~\bibnamefont
  {Kabsch}},\ }\href {\doibase 10.1107/S0907444909047337} {\bibfield  {journal}
  {\bibinfo  {journal} {{A}cta {C}ryst. {D}}\ }\textbf {\bibinfo {volume}
  {66}},\ \bibinfo {pages} {125} (\bibinfo {year} {2010})}\BibitemShut
  {NoStop}%
\bibitem [{\citenamefont {Verchenko}\ \emph
  {et~al.}(2019{\natexlab{b}})\citenamefont {Verchenko}, \citenamefont
  {Zubtsovskii}, \citenamefont {Wei}, \citenamefont {Tsirlin}, \citenamefont
  {Dikarev},\ and\ \citenamefont {Shevelkov}}]{mogazn}%
  \BibitemOpen
  \bibfield  {author} {\bibinfo {author} {\bibfnamefont {V.~Y.}\ \bibnamefont
  {Verchenko}}, \bibinfo {author} {\bibfnamefont {A.~O.}\ \bibnamefont
  {Zubtsovskii}}, \bibinfo {author} {\bibfnamefont {Z.}~\bibnamefont {Wei}},
  \bibinfo {author} {\bibfnamefont {A.~A.}\ \bibnamefont {Tsirlin}}, \bibinfo
  {author} {\bibfnamefont {E.~V.}\ \bibnamefont {Dikarev}}, \ and\ \bibinfo
  {author} {\bibfnamefont {A.~V.}\ \bibnamefont {Shevelkov}},\ }\href {\doibase
  10.1039/c8dt04982c} {\bibfield  {journal} {\bibinfo  {journal} {{D}alton
  {T}rans.}\ }\textbf {\bibinfo {volume} {48}},\ \bibinfo {pages} {7853}
  (\bibinfo {year} {2019}{\natexlab{b}})}\BibitemShut {NoStop}%
\bibitem [{\citenamefont {Poole}(2000)}]{proxy}%
  \BibitemOpen
  \bibfield  {author} {\bibinfo {author} {\bibfnamefont {C.~P.}\ \bibnamefont
  {Poole}},\ }\href@noop {} {\emph {\bibinfo {title} {{H}andbook of
  {S}uperconductivity}}}\ (\bibinfo  {publisher} {{A}cademic {P}ress},\
  \bibinfo {address} {525 B Street, Suite 1900, San Diego, CA 92101-4495,
  USA},\ \bibinfo {year} {2000})\BibitemShut {NoStop}%
\bibitem [{\citenamefont {Padamsee}\ \emph {et~al.}(1973)\citenamefont
  {Padamsee}, \citenamefont {Neighbor},\ and\ \citenamefont
  {Shiffman}}]{alpha-1}%
  \BibitemOpen
  \bibfield  {author} {\bibinfo {author} {\bibfnamefont {H.}~\bibnamefont
  {Padamsee}}, \bibinfo {author} {\bibfnamefont {J.~E.}\ \bibnamefont
  {Neighbor}}, \ and\ \bibinfo {author} {\bibfnamefont {C.~A.}\ \bibnamefont
  {Shiffman}},\ }\href {\doibase 10.1007/BF00654872} {\bibfield  {journal}
  {\bibinfo  {journal} {{J}. {L}ow {T}emp. {P}hys.}\ }\textbf {\bibinfo
  {volume} {12}},\ \bibinfo {pages} {387} (\bibinfo {year} {1973})}\BibitemShut
  {NoStop}%
\bibitem [{\citenamefont {Johnston}(2013)}]{alpha-2}%
  \BibitemOpen
  \bibfield  {author} {\bibinfo {author} {\bibfnamefont {D.~C.}\ \bibnamefont
  {Johnston}},\ }\href {\doibase 10.1088/0953-2048/26/11/115011} {\bibfield
  {journal} {\bibinfo  {journal} {{S}upercond. {S}ci. {T}echnol.}\ }\textbf
  {\bibinfo {volume} {26}},\ \bibinfo {pages} {115011} (\bibinfo {year}
  {2013})}\BibitemShut {NoStop}%
\end{thebibliography}%
\end{document}